# Wireless Communication Performance Testing: From Laboratory Environment to Research Vessel

Andrei-Raoul Morariu, Andreas Strandberg, Bogdan Iancu, Jerker Bjorkqvist

*Faculty of Science and Engineering*
*Åbo Akademi University*
*Vesilinnantie 3, 20500 Turku, Finland*

firstname.lastname@abo.fi

*Abstract* - **This study investigates signal transmission within a shared spectrum, focusing on measurements conducted both in laboratory and outdoor environments. The objective was to demonstrate how laboratory objects obstructing the line of sight can attenuate the signal between a transmitter (Tx) and a receiver (Rx). Additionally, we examined the impact of distance and placement in various locations aboard an electric research boat on signal transmission efficiency. These findings contribute to understanding whether the environmental factors influence wireless communication in dynamic and obstructed environments.**

## I. Introduction

Wireless channel performance is critical for efficient wireless communication. It is based on factors such as signal strength, signal-to-noise ratio, bandwidth, and latency. Bandwidth and throughput, representing data transfer capacity and the rate of successful data transfer, are both impacted by interference. Channel reliability is affected by multipath fading, where the signal is reflected off surfaces, and error rates, including both bit and packet error. In this article, we observe the radio spectrum to identify interference around our target signal and examine the impact of various materials in a laboratory environment and on an electric research vessel on signal strength.

The expansion of wireless systems and the accelerated development of radio communication technologies have intensified the need for smarter spectrum management [1]. This pressure necessitates that the spectrum be utilised efficiently, economically, and fairly. Emerging technologies such as broadband, 5G and 5GA, which require more spectrum for growth, further increase the complexity. In the case of 6G, the RF spectrum may extend into the *milimiterWaves* range and potential exceed 100 THZ [2]. With the rapid growth of wireless systems and the fast-paced evolution of radio communication technologies, there has been increasing pressure on RF spectrum management to ensure that the spectrum is used efficiently, economically, and equitably [3]. Emerging technologies, such as broadband, 5G and the upcoming 6G [4], require additional spectrum to sustain their rapid growth, further intensifying the demand.

All sectors—government, private industries, and telecommunications providers—are experiencing demands for spectrum resources across applications. However, the issue lies not in a shortage of spectrum, but in how it is utilised. Studies show that in certain locations or at specific times, up to 70\% [3] of the allocated spectrum may remain idle despite being officially reserved. This suggests that while the radio spectrum is abundant, the real challenge is the lack of accessible and affordable communication infrastructure.

Water reflection causes multipath fading in wireless communications in a maritime environment. On water there are no obstacles between the two points, and the signal will travel directly in the line of sight and through reflections in the water. Raulefs et al. [5] examine the signals loss path of distance and their findings reveal that using different signal frequencies can be an alternative to using separate antennas at different heights to minimize the path loss.

Lin et al. [6] examine how surface wave dynamics impact water-to-air optical wireless communication performance. Their findings reveal that small-scale waves cause rapid fluctuations in signal strength, increasing bit error rates, while large-scale waves lead to gradual signal fading by altering the light spot's coverage.

The global management and regulation of the RF spectrum are governed by the World Radiocommunication Conferences (WRC), a supreme body responsible for revising the Radio Regulations. WRC decisions are informed by national studies and reports produced by ITU-R Study Groups [7]. These Study Groups play a role in developing International Telecommunication Union - Radiocommunication Sector (ITU-R) recommendations regarding the technical characteristics and operational procedures for radio communication services and systems. Additionally, they draft the technical foundations for radio communication conferences and compile handbooks on spectrum management and emerging radio technologies. Conference Preparatory Meetings (CPM) prepare consolidated reports that cover the technical, operational, regulatory, and procedural bases for a WRC. They synthesize the output from the study groups and the special committees, along with any newly submitted materials, ensuring that each conference is well-informed and can effectively address the growing challenges of spectrum management.

The frequency spectrum in Finland showcased in Figure 1 is evenly distributed among the three telecom operators—DNA, Elisa, and Telia—across various bands, ensuring balanced access to uplink and downlink frequencies. This strategic allocation across the 700 MHz, 800 MHz, 900 MHz, 1800 MHz, and 2100 MHz bands allows each operator to offer robust mobile communication services, including LTE and 5G, across the country. In the 2600 MHz (n38) frequency band, Elisa is the only operator holding a spectrum from 2570 to 2620 MHz.

In this context, spectrum analysis becomes critical for identifying optimal frequencies for reliable communication, minimising interference from environmental factors like water and cloud reflections and ensuring reliable communication for ships at sea [8].

The paper is structured as follows. Section I presents the motivators of our study. The results of the laboratory and outside spectrum measurements are presented in Section II. In Section III, we discuss the Ship Area Network and the standard that makes integration it possible. Section IV concludes the article.

## II. RESULTS/EXPERIMENTS

### A. Signal Transmission: Laboratory Environment

The laboratory environment was a home lab environment in a shared space in our offices. We measured that there was no traffic on the frequency with which we tested our equipment. Our goal with our lab tests was to observe how common household items affect the signal strength of a 5G generated signal.

In Table I, we compared the signals with a recording taken with the transmitter and receiver two meters apart and at eye level. We have marked it **1** for being similar to the original, **2** for being clearer, and **0** for a worse signal. As observed, where the transmitter was placed inside an object an amplification of the signal was observed, and in the other cases where the receiver was placed inside objects, the signal was unreadable from the background noise.

TABLE I
LABORATORY TESTS ON SIGNAL TRANSMISSION

| Lab test | Component placement | |
|---|---|---|
| Location | Transmitter | Receiver |
| Microwave 3.4 GHz, inside | 1 | 2 |
| Microwave 2.4 GHz, inside | 2 | 0 |
| Ceramic pot, inside | 2 | 0 |
| Cable's max length of 4 m | 0 | 0 |
| Behind a fire hydrant | 1 | 1 |
| Behind a concrete pillar | 2 | 2 |
| Aluminium kettle | 2 | 0 |
| Wi-Fi mesh station | 1 | 1 |
| Ceramic pot 2 m distance, inside | 2 | 0 |
| Ceramic pot 2 m distance, face away, inside | 1 | 0 |

### B. Signal Transmission: Electric Research Boat

The testing on the electric boat eM/S Salama [10] was to observe any differences in signal strength in the radio frequency spectrum with a transmitting IoT device being moved into different hatches and locations on the vessel, as highlighted in Fig. 2. In the highlighted area of Figure 1, it can be observed that the generated signal fits into the spectrum allocation of two network operators in Finland. The test was finished by recording the spectrum while the vessel drove out from the harbour 500 meters into the open water. With the graph in Fig. 4 we have illustrated the drop in the captured signal strength of our IoT device. It took about a minute for the vessel to exit the docking spot, and from minute 1, there is an almost even decline in the strength by 5 dBm per minute. Before the vessel had reached the designated stop point, the signal was lost in the signal noise floor of our recording.

For our test cases of different locations and enclosed spaces, we observed minimal to no differences in the signal strength of the IoT board. In Table II we have listed all test cases performed during our test with the average signal strength per test case. Test case 5 is the reference for the power registered, having the Tx and Rx near each other. Also, the test cases 6-8 are not visualised in Fig. 2, given that no power was recorded. In the screenshot, Fig. 3 we observed the signal during our tests.

TABLE II
RESEARCH VESSEL TEST CASES AND DESCRIPTION

| Test case | Description | Power (dBm) |
|---|---|---|
| 1 | Device in glass cabin inside vessel | -65 |
| 2 | Device in motor hatch, metal/plastic covered | -68 |
| 3 | Device on roof of vessel's cabin, eye of sight | -73 |
| 4 | Device in aluminium hatch behind cabin | -68 |
| 5 | Device held in front of antenna on shore | -70 |
| 6 | Vessel driving away, placement test case | - |
| 7 | Vessel out on water, placement test case | - |
| 8 | Device not transmitting, background signal traffic | - |

## III. DISCUSSION

A customary Ship Area Network (SAN) provides a framework for remote management and control of sensors and instruments on board ships, facilitating tasks such as data collection, system monitoring and control that are critical to maritime safety. These processes cover a variety of shipboard locations, such as the administrative, bridge, and engine rooms, in addition to remote activities outside the ship. According to international standards, SANs are usually arranged in a hierarchical three-tier structure: Instrument layer, Process layer, and Integrated Ship Control (ISC) layer [11]. These networks are essential for managing safety-critical data, supervising on-board devices, and enabling navigation. In the past, SANs have relied on wired communication, using protocols such as Controller Area Networks (CAN), Ethernet cables, and analog signals with a current of 4 to 20 mA.

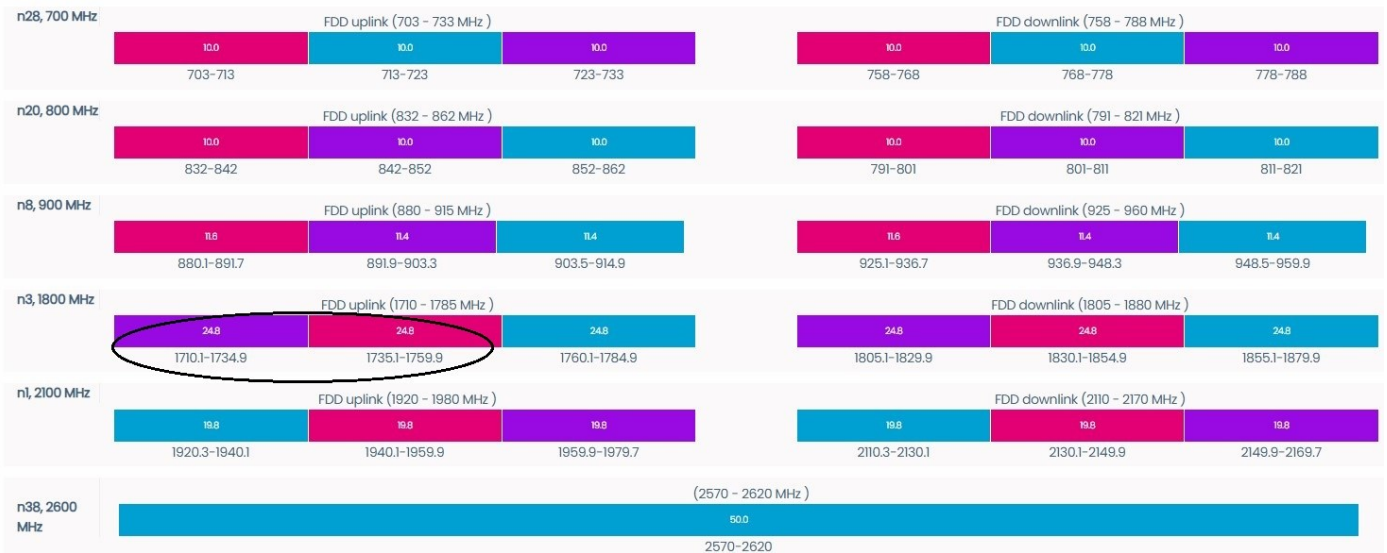

Fig. 1 Finland spectrum of three network operators within the range of 700 MHz-2600MHz, as depicted in [9].

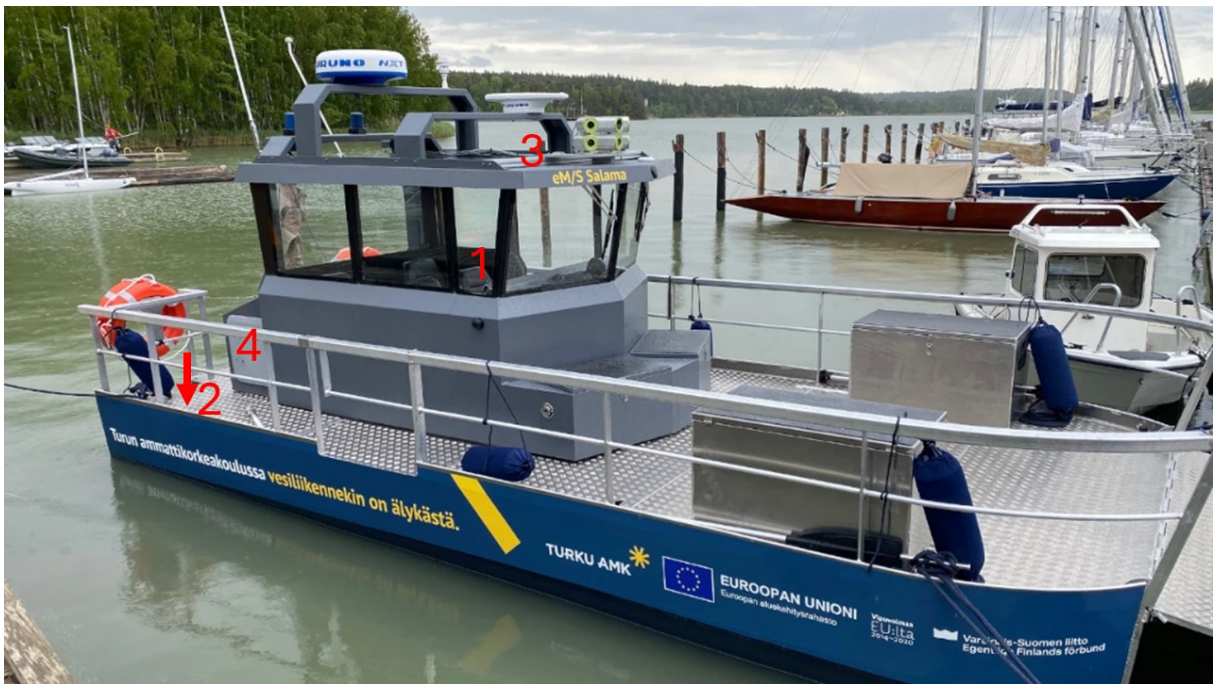

Fig. 2 Electric boat eM/S Salama [10] and the Tx placements as on Table II

The adoption of the National Marine Electronics Association (NMEA) 2000 standard made possible tighter integration of SANs with navigation systems [12]. Using technologies such as WiMedia, wireless communication, albeit restricted to particular compartments, has been suggested for interactions between devices in the ISC and Instrument layers. The idea of hybrid (wired and wireless) networks has become popular in ship applications recently, mirroring developments in other sectors. Access to previously inaccessible data is improved through wireless sensor networks, which are made up of dispersed devices that measure environmental factors such as temperature, pressure, or vibration. Technologies such as Bluetooth, Wi-Fi, ZigBee, and GSM are used by these networks to offer location-independent flexible communication. Standardized data formats, radio signals, and a strong network architecture are all required for efficient wireless data transport.

Although these components operate separately, they must be in line for smooth wireless operations. The development of wireless technologies has made it possible for industrial automation and marine systems to be adopted more widely, signalling improvements in automation and operational effectiveness on ships.

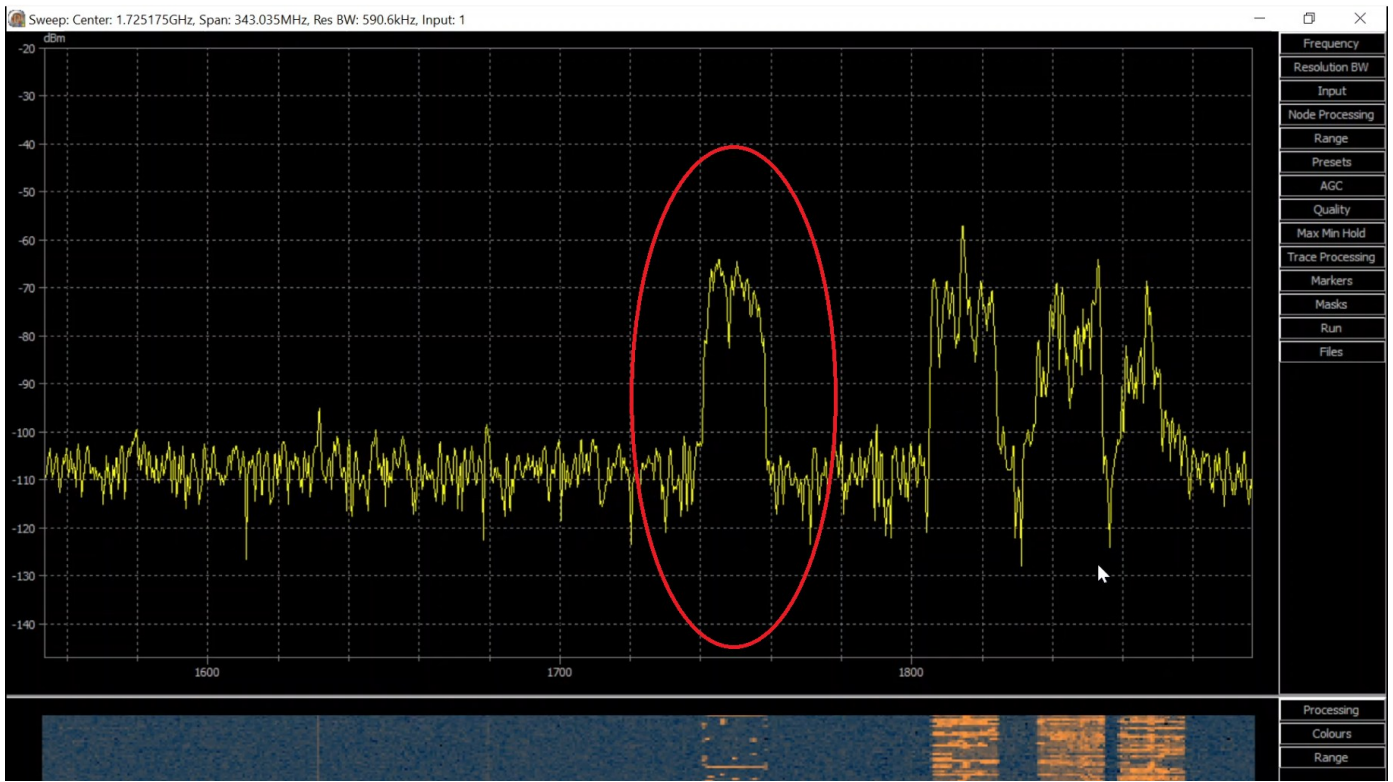

Fig. 3 Test Case 4. Tx was placed inside the hatch on the back of the cabin

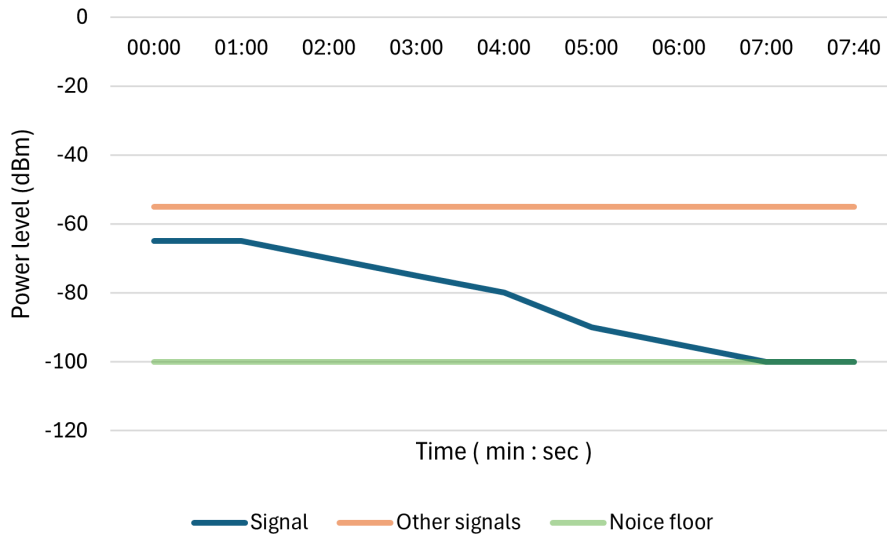

Fig. 4 Signal strength recorded when the electric boat moved from shore to 500m distance

Wireless communication quality can be affected by the main engine, other machinery, the structural walls of the ship compartment, and even passenger actions such as opening and closing metallic doors. Compared to typical environments, be it indoor or outdoor, the ship's environment, which is defined by a complex metallic framework, presents more difficulties for wireless signal transmission [13].

Masnicki et al. describe in [13] their efforts to enhance signal transmission quality in a case study focusing on two key areas: addressing the impact of shipboard partitions and selecting an optimal wireless network configuration. Their detrimental effects on radio signal strength and quality are successfully reduced by placing repeater nodes in designated tunnels behind partition walls, such as bulkheads and watertight doors. When it comes to ZigBee networks, they emphasise the self-organization features and ZigBee routers — two essential components for setting up and maximizing onboard networks. Wi-Fi was primarily used for communication within wireless LANs and multimedia access through Access Point devices, which facilitated internet-based activities like website browsing, database access, and email management. The development of distributed measurement and control systems customised for shipboard settings has been made possible by the availability of high-performance Wi-Fi repeaters, which have improved network capabilities [13].

## IV. Conclusions

The findings of this study reveal that spectrum measurements of a signal between a transmitter and receiver in a laboratory setting showed minimal interference from obstacles like a concrete pillar or a fire hydrant. A similar experiment conducted on an electric boat indicated little to no difference in transmission signal strength when the IoT board was placed in various locations on the boat. Additionally, as the boat moved 500 meters away from the shore, a typical reduction in signal power was observed due to the distance. Overall, this study demonstrates that materials in both laboratory and electric boat environments had minimal to no effect on the signal strength transmitted from the IoT board.

## Acknowledgments

This work was supported by the Project 5G Advanced. The authors express their gratitude for the financial support that made this research possible.